\documentclass[aps,prl,preprint,groupedaddress]{revtex4}

\usepackage{bm} 
\usepackage{amsmath}
\usepackage{amsfonts}
\usepackage{amssymb} 
\usepackage{amsbsy}
\usepackage[dvips]{epsfig,color} 
\usepackage{graphicx}
\usepackage{psfrag}
\usepackage{wasysym}
\usepackage{latexsym}

\newcommand{\dashedline}{\protect\rule[2pt]{4pt}{1pt}\,\,\protect\rule[2pt]{4pt}{1pt}\,\,\protect\rule[2pt]{4pt}{1pt}}

\newcommand{\fullline}{\protect\rule[2pt]{15pt}{1pt}}
\newcommand{\dasheddottedline}{\protect\rule[2pt]{4pt}{1pt} \!$\cdot$ \!\protect\rule[2pt]{4pt}{1pt} \!$\cdot$}

\begin{document}
\title{Nonequilibrium domain formation by pressure fluctuations} 
\author{Marco G. Mazza$^{\dagger}$ and Martin Schoen$^{\dagger,\ddagger}$} 

\affiliation{
  $^{\dagger}$Stranski-Laboratorium f\"ur Physikalische und
  Theoretische Chemie, Technische Universit\"at Berlin, Stra{\ss}e des
  17. Juni 135, 10623 Berlin, Germany\\
  $^{\ddagger}$Department of Chemical and Biomolecular Engineering, North
  Carolina State University, 911 Partners Way, Raleigh, NC 27695,  U.S.A.}

\date{\today}

\maketitle

{\bf Fluctuations in thermal many-particle systems reflect fundamental
  dynamical processes in both equilibrium and nonequilibrium (NEQ) physics.
  In NEQ systems \cite{ritort} fluctuations are important in a variety of
  contexts ranging from pattern formation \cite{hohenberg,vdbroeck} to
  molecular motors \cite{schaller,kierfeld,narayan,prost}. Here, we address
  the question if and how fluctuations may be employed to characterize and
  control pattern formation in NEQ nanoscopic systems. We report computer
  simulations of a liquid crystal system of prolate molecules (mesogens)
  sandwiched between flat walls, and exposed to a time-dependent external
  field. We find that a switchable smectic domain forms for sufficiently high
  frequency.  Although pressure and temperature are too low to induce an
  equilibrium smectic phase, the fluctuations of the pressure in the NEQ
  steady state match the pressure fluctuations characteristic of the
  equilibrium smectic phase.  Furthermore, the {\it wall-normal} pressure
  fluctuations give rise to a {\it tangential} ``fluctuation-vorticity''
  tensor that specifies the symmetry-breaking direction of the smectic layers.
  Our calculations demonstrate a novel method through which nanomaterials with
  a high degree of molecular order may be manufactured in principle.}

In nature, systems out of equilibrium are the rule and not the exception
\cite{racz}. Yet, only recently, with the growing interest in nano- and
mesoscopic phenomena, NEQ thermodynamics has attracted a remarkably growing
interest \cite{ritort,seifert08}. Some profound and pioneering results are
already available \cite{evans93,galla95,jarzin,seifert05}, but a coherent
picture of NEQ physics is still lacking.

A defining difference between the physics of NEQ and equilibrium systems is
the presence of nonvanishing currents in the former, which are maintained by
mechanical, thermal or chemical driving forces\cite{hohenberg,schmiedl}. In
fact, many works \cite{zia,hurtado_pre10,hurtado_pnas11,bertini,derrida} have
addressed the statistical properties of these currents. Nonetheless, their
physical role in a fluid is still not clear.

Here, we perform molecular dynamics simulations of $N=4000$ Gay-Berne-Kihara
(GBK) molecules in presence of two atomically smooth flat walls. The GBK model
has been successfully used to reproduce a typical equilibrium liquid crystal
(LC) phase diagram for prolate mesogens \cite{martinez05}. The
isobaric-isothermal ensemble was used to avoid unphysical stresses on the
system due to the combination of a cubic simulation box and the formation of
anisotropic phases.

The interaction of the molecules with the walls is modeled with a
Lennard-Jones potential. Because of the presence of the walls the pressure in
the system depends on the direction, that is, it is a second-rank tensor, and
not a scalar
\begin{equation}
\boldsymbol{\mathcal{P}}\equiv\frac{1}{V}\sum_{i=1}^N\Big[m\mathbf{v}_i\otimes\mathbf{v}_i+\mathbf{r}_{i}\otimes\mathbf{f}_{i}\Big]
\end{equation}
where $\mathbf{v}_i$ is the velocity of the $i$th particle, $m$ its mass,
$\mathbf{r}_i$ its position vector, $\mathbf{f}_i$ the total force acting on
it, $V$ the total volume of the system, and the operator $\otimes$ represents
the dyadic product. Further, because of the planar geometry of the system,
only the diagonal Cartesian components $\mathcal{P}_{xx}, \mathcal{P}_{yy}$
and $\mathcal{P}_{zz}$ are nonzero, and the hydrostatic pressure is
$P\equiv\langle\mathcal{P}_{xx}\rangle=\langle\mathcal{P}_{yy}\rangle$ (see
Methods).

In the case of anisotropic molecules, it is important to specify their
preferential alignment at the walls. A suitable quantity is the so-called
``anchoring function'' $g(\hat{\mathbf{u}})$ which discriminates energetically
the orientation $\hat{\mathbf{u}}$ of a LC molecule with respect to a surface
\cite{sonin}, effectively defining a preferential direction, also called
``easy axis'' \cite{abbott11}. We consider a system confined by walls whose
surface properties change periodically with time.  Specifically, we give a
temporal dependence to the anchoring function $g(\hat{\mathbf{u}},t)\equiv
A\,(u_xV_x+u_yV_y+u_zV_z)$, where $V_x=V_y=\sin(\omega t)$, $V_z=\cos(\omega
t)$, $A$ is a constant, $\omega$ is the angular frequency of the sinusoidal
external field, and $t$ is time. The effect of the external field is then to
rotate the walls' easy axes with time. The two easy axes rotate in phase.
Time-dependent, responsive surfaces are a growing field of research
\cite{stuart}.  Clare \emph{et al}. \cite{clare06} have demonstrated that any
specific anchoring of LCs can be selectively obtained by grafting
semifluorinated organosilanes onto a surface. Further, the realization of time
varying decorated surfaces has been reported where variable pH
\cite{ionov04,matthews}, electrochemical properties \cite{song}, or UV-light
irradiation are used as agents to effect the time dependence \cite{feng}.

\begin{figure}
\begin{center}
\includegraphics[scale=0.3]{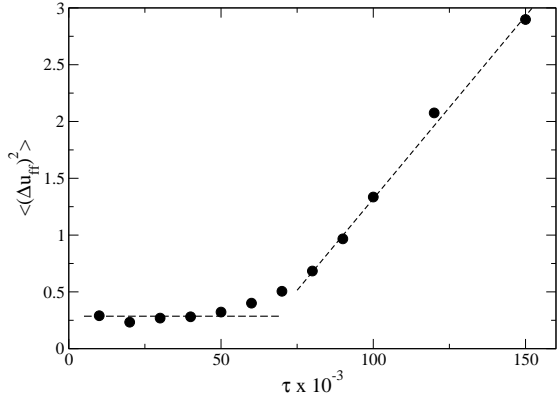}
\end{center}
\caption{{\bf Crossover of the energy fluctuations upon formation of a NESS.}
  Dependence of the fluid-fluid energy fluctuations on the period of the
  external field $\tau$. The dashed lines represent linear fits for large and
  small values of $\tau$.}\label{ener-fluct}
\end{figure}

We present results for simulations in a range of temperature $T=4.0-6.0$ at
fixed $P=3.6$ and investigate the behavior of the system as the period of the
external field $\tau=2\pi/\omega$ is varied.  We use dimensionless units
throughout this work (see Methods). We first analyze the dependence of the
energy fluctuations of the fluid on $\tau$.  Figure \ref{ener-fluct} shows
that $\langle(\Delta u_{ff})^2\rangle$ grows linearly with $\tau$ at large
$\tau$ (which is the usual behavior of the energy fluctuations for a
sinusoidal field). At $\tau\approx 7\times10^4$ there is a crossover to a
plateau.  Because $\langle(\Delta u_{ff})^2\rangle$ is proportional to the
specific heat of the fluid, this crossover reflects a structural rearrangement
of the system occurring at small periods (high frequency).

\begin{figure}
\includegraphics[scale=0.9]{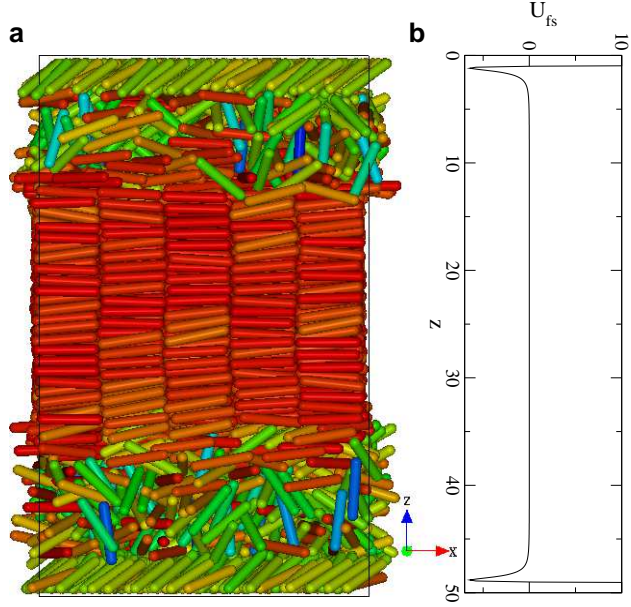}
\caption{{\bf Molecular configuration of the simulated fluid and comparison
    with the extension of the fluid-substrate interaction.} {\bf a}, Snapshot
  showing the lateral view of the system in the $xz$-plane at $P=3.6$, $T=4.0$
  and $\tau=3\times10^4$.  The molecules closest to the walls (top and bottom
  layers) rotate with the external field. A smectic domain is clearly visible
  in the center of the system. {\bf b}, Plot of the fluid-substrate
  interaction $U_{fs}$ (see Methods) showing that it is effectively different
  from zero only in a region less than $5$ molecular diameter in
  size.}\label{snap}
\end{figure}

Visual inspection of the molecular configurations reveals (Fig. \ref{snap}a)
that the molecules in the central portion of the fluid assemble in a well
defined smectic domain (SD). This domain develops after a relatively small
number of cycles of the external field and persists as long as the field is
switched on; the SD disappears quickly once the external field is switched
off.  The fluid becomes heterogeneous. The molecules directly in contact with
the walls rotate as the anchoring associated with the field changes.  Between
the contact layer at the walls and the SD there is a very turbulent layer,
which does not show any spatial nor orientational order.  Close observation of
the molecular dynamics shows that individual molecules constantly leave this
turbulent layer to join the SD or vice versa; however, the SD remains a stable
feature. This is an instance of a nonequilibrium steady state (NESS). Only the
molecules in the contact layer (i.e, the layer closest to the wall) are
subject to direct interaction with the wall because of the short-range nature
of the fluid-substrate potential $U_{fs}$ (see Fig.  \ref{snap}b and Methods).
Therefore, the formation of SD must not be confused with a confinement effect,
because the SD forms in the region where $U_{fs}\approx0$. Hence, we believe
the formation of the SD to be a general consequence of a time-dependent
external field, irrespective of the precise realization of this field, so that
our results are relevant to a broad class of physical situations.

\begin{figure}
\begin{center}
\includegraphics[scale=0.33]{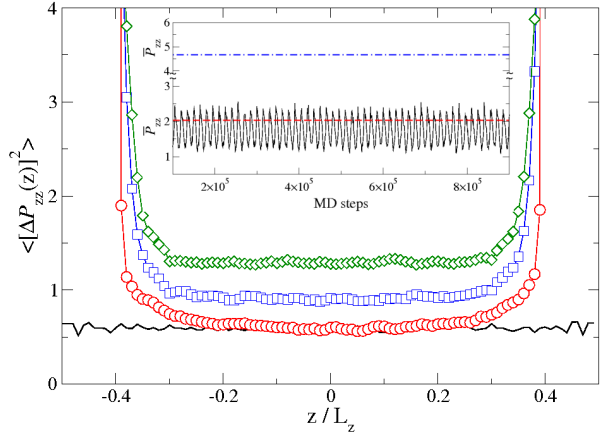}
\end{center}
\caption{{\bf Normal pressure fluctuations.} Dependence of the pressure tensor
  fluctuations on position along the $z$-axis for NEQ simulations at $P=3.6$,
  $T=4.0$ for $\tau=3\times 10^4$ ($\bigcirc$), $\tau=7\times 10^4$
  ($\square$), and $\tau=9\times 10^4$ ($\Diamond$).  The black line shows the
  value of the pressure tensor fluctuations for an equilibrium simulation in
  the smectic phase. Inset shows the local pressure integrated over the SD
  volume $\overline{\mathcal{P}}_{zz}$ ($\fullline$).  The NEQ value
  oscillates with time following the external field, and, interestingly, its
  average value is close to the value characteristic of an {\it equilibrium}
  isotropic phase (\dashedline), which is the thermodynamic equilibrium
  state at this $P$ and $T$.  The equilibrium value characteristic of the
  smectic phase is much larger (\dasheddottedline). From this we conclude that
  the local pressure is not large enough to drive the formation of a
  SD.}\label{press_fluct}
\end{figure}

Now, a natural question to ask is: Why should a SD form?  In equilibrium, the
lowest $P$ for which a smectic phase forms at $T=4.0$ is more than twice the
value of $P$ investigated in this work.  Thus, the equilibrium phase
transition is too far removed to play any role here. It is also important to
note that even the local value of the pressure in the region where the SD
forms is too low to explain a smectic state. The inset in
Fig.~\ref{press_fluct} shows that the local pressure integrated over the SD
volume, $\overline{\mathcal{P}}_{zz}$, is too low compared with the same
quantity but calculated for an equilibrium smectic state at the same $P$ and
$T$.

We then consider the temporal fluctuations of the pressure $\langle[\Delta
\mathcal{P}_{zz}(z)]^2\rangle\equiv\langle[\mathcal{P}_{zz}(z,t)]^2\rangle-\langle[\mathcal{P}_{zz}(z,t)]\rangle^2$.
Figure \ref{press_fluct} shows the dependence of $\langle[\Delta
\mathcal{P}_{zz}(z)]^2\rangle$ on the position along the $z$-axis. The
confining walls are located at $z/L_z=\pm 1/2$, where the pressure
fluctuations are very large due to the molecular rotation.  As we move towards
the center of the system, $\langle[\Delta \mathcal{P}_{zz}(z)]^2\rangle$
decreases rapidly until it reaches an almost constant value. It is the main
observation of this study that when a SD forms the NEQ pressure fluctuations
match the value of the equilibrium pressure fluctuations in a smectic phase
\begin{equation}\label{eqpfluct}
\langle[\Delta\mathcal{P}_{zz}(z)]^2\rangle_{\mathrm{NEQ}}=\langle[\Delta
\mathcal{P}_{zz}(z)]^2\rangle_{\mathrm{EQ}}\,. 
\end{equation}
Also, the region of the plateau of $\langle[\Delta
\mathcal{P}_{zz}(z)]^2\rangle$ coincides with the location of the SD. As
$\tau$ increases the SD shrinks and becomes less coherent; this correlates
very well with the behavior of $\langle[\Delta \mathcal{P}_{zz}(z)]^2\rangle$
in Fig.~\ref{press_fluct} for $\tau\geqslant6\times10^4$. This value deviates
increasingly from $\langle[\Delta
\mathcal{P}_{zz}(z)]^2\rangle_{\mathrm{EQ}}$.


The instantaneous value of $\mathcal{P}_{zz}(z,t)$ may be treated as a
stochastic variable resulting from the chaotic molecular motion and the
oscillatory behavior at the walls.  Therefore, to rationalize the coincidence
between the SD formation and equation (\ref{eqpfluct}) we turn to a
statistical description of the pressure profile in terms of a Fokker-Planck
equation for the probability
$\Pi(p,t)\equiv\langle\delta(\mathcal{P}_{zz}(z,t)-p)\rangle$
\begin{equation}
\frac{\partial\Pi(p,t)}{\partial t}=-\frac{\partial}{\partial p}[C(p)\Pi(p,t)]
+\frac{1}{2}\frac{\partial^2}{\partial p^2}[D(p)\Pi(p,t)]
\end{equation}
where $\delta(x)$ is the Dirac $\delta$-function, $\langle\cdot\cdot\cdot\rangle$
represents the average over the molecular noise, $C(p)\equiv\langle\Delta
p\rangle/\Delta t=\langle\Delta\mathcal{P}_{zz}\rangle/\Delta t$,
$D(z)\equiv\langle[\Delta p]^2\rangle/\Delta
t=\langle[\Delta\mathcal{P}_{zz}]^2\rangle/\Delta t$ in the limit $\Delta t
\to 0$ \cite{vankampen}.  Now, it is readily seen that $C(p)$ vanishes because
the field configuration is symmetric and therefore the transition probability
is symmetric in the increment $\Delta\mathcal{P}_{zz}$.  Hence, the
probability $\Pi(p,t)$ is governed not by the value of $\mathcal{P}_{zz}$ but
rather by the fluctuations similar to ordinary Brownian motion.

\begin{figure}
\begin{center}
\includegraphics[scale=0.8]{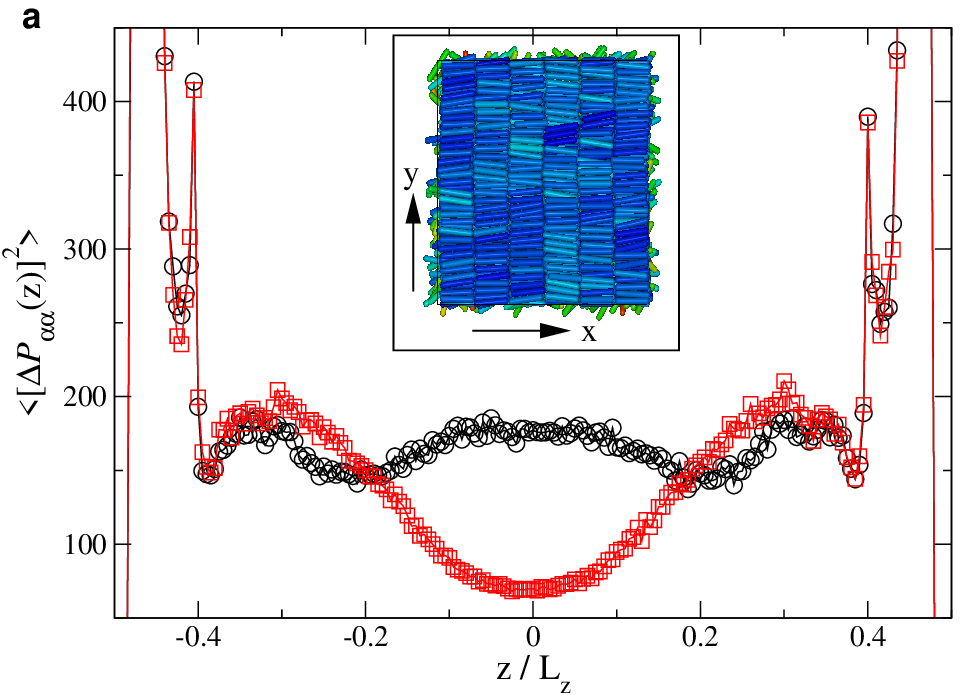}
\includegraphics[scale=0.8]{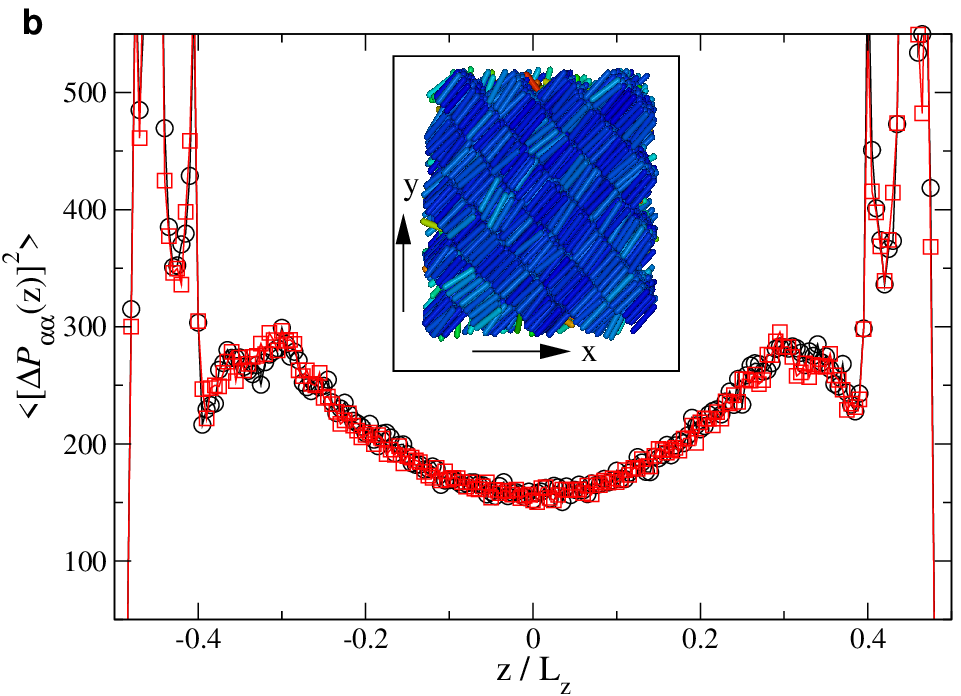}
\end{center}
\caption{{\bf Tangential pressure fluctuations.} Dependence of the pressure
  tensor fluctuations $\langle[\Delta \mathcal{P}_{xx}(z)]^2\rangle$ (black)
  and $\langle[\Delta \mathcal{P}_{yy}(z)]^2\rangle$ (red) on position along
  the $z$-axis for NEQ simulations at $P=3.6$, $T=4.0$ and $\tau=3\times10^4$
  for a SD configuration with the layer-normal parallel to the $x$-axis ({\bf
    a}), and at $45^\circ$ with the $x$-axis ({\bf b}).  Insets show the top
  view of the cross sections of the two systems with the SD corresponding to
  the tangential pressure fluctuations calculated here.}\label{press_fluct_yy}
\end{figure}

Because the SD is a NESS we assume that mechanical stability is locally valid
in the central part of the fluid, sufficiently removed from the walls. Locally
then,
\begin{equation}\label{nablap}
\nabla\cdot\boldsymbol{\mathcal{P}}={\bf 0}.
\end{equation}
In equilibrium, from equation (\ref{nablap}) follows that
$\mathcal{P}_{zz}(z)=const.$ in the entire fluid. From the fact that
$\langle[\Delta \mathcal{P}_{zz}(z)]^2\rangle$ does not depend on $z$ in the
central portion of the fluid (where the SD forms, see Fig. \ref{press_fluct})
we are led to assume that a similar relation to equation (\ref{nablap}) is valid
for the pressure fluctuations
\begin{equation}\label{nablafluctp}
\nabla\cdot(\Delta\boldsymbol{\mathcal{P}})={\bf 0}.
\end{equation}

Similar to standard hydrodynamics, we can then define a
``fluctuation-vorticity'' tensor associated to the pressure fluctuations,
$\boldsymbol{\omega}\equiv\nabla\times\Delta\boldsymbol{\mathcal{P}}$. Because
of the planar geometry of our system the pressure tensor components are only
functions of $z$, such that
$\omega_{xx}=-\partial(\Delta\mathcal{P}_{yy})/\partial z$ and
$\omega_{yy}=\partial(\Delta\mathcal{P}_{xx})/\partial z$. A larger slope of
$\Delta\mathcal{P}_{xx}$ or $\Delta\mathcal{P}_{yy}$ then implies a larger
$\omega_{yy}$ or $\omega_{xx}$, respectively. To test whether
$\boldsymbol{\omega}$ has physical significance we consider two systems with
the same normal pressure fluctuations, i.e. the same $\tau$, but with
different smectic-layer normals. Figure \ref{press_fluct_yy}a shows the
pressure-fluctuation profile $\langle[\Delta
\mathcal{P}_{\alpha\alpha}(z)]^2\rangle$, $\alpha=x,y$, for a system
exhibiting a SD with a smectic-layer normal  parallel to the $x$-axis.
The SD extends in the region $|z|/L_z\lesssim0.2$.  In the same region
$\Delta\mathcal{P}_{xx}$ has zero slope, while $\Delta\mathcal{P}_{yy}$
exhibits a large slope. This, in turn, implies a vanishing $\omega_{yy}$ and a
large $\omega_{xx}$. The relative magnitude of $\omega_{xx}$ and $\omega_{yy}$
correlates with the orientation of the smectic-layer normal. Further, in Fig.
\ref{press_fluct_yy}b we show the pressure fluctuation profile for the second
SD whose smectic-layer normal is at an angle of $45^\circ$ with the $x$-axis.
The two curves coincide (within numerical accuracy) indicating equal
tangential components of $\boldsymbol{\omega}$.  Therefore, from Fig.
\ref{press_fluct_yy} we conclude that the fluctuation-vorticity
$\boldsymbol{\omega}$ determines the symmetry breaking direction of alignment
of the SD.

To conclude, we find evidence from NEQ computer simulations that pressure
fluctuations can be easily tuned to drive a fluid system to a
far-from-equilibrium state. The role of current fluctuations has been
recognized \cite{zia,hurtado_pnas11} as a stochastic variable characterizing
NESS. Here, the physical picture emerging is that fluctuations in the momentum
current (i) determine the NESS, and (ii) give rise to a secondary field that
breaks the rotational symmetry in the $xy$-plane.

Self-assembly of molecules or supramolecular particles into layers, membranes,
and vesicles is revolutionizing our control of matter across multiple length
scales with far-reaching applications in nanofluidic devices
\cite{brake,woltman,abbott11}. Chemico-physical properties are carefully tuned
to obtain the desired features \cite{Genzer08}. However, in most cases they do
not have any temporal dependence. The richness of NEQ phenomena in simple
systems may suggest that combining the powerful new techniques of nanoscopic
control with the application of time dependent external fields (temperature,
electric or magnetic field, pressure and pH) may usher new ways to induce
molecular self-assembly and even to simplify known tasks. In particular, the
vorticity field may be used in the future to control the orientation of the
ordered smectic domains which could be useful to manufacture new nanoscopic
materials with distinct materials properties.

\acknowledgments Financial support from the International Graduate Research
Training Group 1524 is gratefully acknowledged.

\section{Methods}

The fluid-substrate interaction is modeled with an ``integrated''
Lennard-Jones potential
\begin{equation}
U_{fs}=4\epsilon_{fs}\rho_{\rm s}\left[\left(\frac{\sigma}{d^{\rm
        m}_{ik}}\right)^{10}
-\left(\frac{\sigma}{d^{\rm m}_{ik}}\right)^4g(\hat{\mathbf{u}},t)\right]
\end{equation}
where $\epsilon_{fs}=1$ and $\rho_{\rm s}\sigma^2=2^{2/3}\pi$ is the areal
density of a single layer of atoms arranged according to the ($100$) plane of
a face-centered cubic lattice. The diameter $\sigma$ of the substrate atoms is
equal to the LC molecular diameter. The quantity $d^{\rm m}_{ik}$ is the
minimum distance \cite{vega94} between a LC molecule and the substrate located
at $z=-L_{\rm z}/2$ ($k=1$) and $z=+L_{\rm z}/2$ ($k=2$). The time-dependent
anchoring $g(\hat{\mathbf{u}},t)$ is included in the attractive part of the
fluid-substrate interaction. Dimensionless units are used throughout, that is,
length is expressed in units of $\sigma$, temperature in units of
$\epsilon_{\mathrm{ff}}/k_{\mathrm{B}}$, time in units of
$(\sigma^2m/\epsilon_{\mathrm{ff}})^{1/2}$ using $m=1$, and pressure $P$ in
units of $\sigma^3/\epsilon_{\mathrm{ff}}$, where $\epsilon_{\mathrm{ff}}$ is
the fluid-fluid interaction energy scale of the GBK model~\cite{martinez05}.

We use a velocity-Verlet algorithm for linear molecules \cite{ilnytskyi02},
and the simulations are carried out in the NPT ensemble using a Nos\'e-Hoover
thermostat \cite{nose,hoover} and an anisotropic Hoover barostat
\cite{ilnytskyi07}, whereby $L_z$ is kept fixed, while $L_x$ and $L_y$ are
allowed to vary independently from each other, resulting in equal lateral
average values of the pressure tensor
$\langle\mathcal{P}_{xx}\rangle=\langle\mathcal{P}_{yy}\rangle$.

We use the ``method of planes'' \cite{todd} to compute the component
$\mathcal{P}_{zz}$ of the pressure tensor. Unfortunately, this method cannot
provide $\mathcal{P}_{xx}$ and $\mathcal{P}_{yy}$ by construction, which
instead we compute following Harasima's method
\cite{harasima}.

\bibliographystyle{prsty.bst}

\end{document}